\documentclass[a4paper,usenatbib]{jpconf}
\usepackage{graphicx}
\usepackage[dvipsnames]{color}
\usepackage{url}
\usepackage{amsmath} 
\usepackage{graphicx}
\usepackage{amssymb}
\usepackage{floatpag}
\usepackage{float}
\usepackage[section]{placeins}
\usepackage{relsize}
\usepackage{url}

\newcommand{\prd}{Physics Review D}

\newcommand{\apj}{ApJ}
\newcommand{\mnras}{MNRAS}
\newcommand{\aap}{A\&A}

\def \prd {PRD}
\def \apj {ApJ}
\def \apjs {ApJS}
\def \apjl {ApJL}
\def \mnras {MNRAS}
\def \aap {A\&A}

\def \pasj {PASJ}
\def \pasa {PASA}

\def \nat {Nature}

\begin{document}

\title{Fast response electromagnetic follow-ups from low latency GW triggers}

\author{ E J Howell$^{1}$, Q Chu$^{1}$, A Rowlinson$^{2}$, H Gao$^{3}$, B Zhang$^{4}$, S J Tingay$^{5,6}$, M Bo\"er$^{7}$ and L Wen$^{1}$  }
\address{$^{1}$School of Physics, University of Western Australia, Crawley WA 6009, Australia}
\address{$^{2}$CSIRO Astronomy and Space Science, Sydney, Australia }
\address{$^{3}$Department of Astronomy, Beijing Normal University, Beijing 100875, China}
\address{$^{4}$Department of Physics and Astronomy, University of Nevada Las Vegas, NV 89154, USA}
\address{$^{5}$International Centre for Radio Astronomy Research, Curtin University, Perth, WA 6845, Australia}
\address{$^{6}$ARC Centre of Excellence for All-sky Astrophysics (CAASTRO)}
\address{$^{7}$CNRS - ARTEMIS, boulevard de l'Observatoire, CS 34229, 06304 Nice Cedex 04, France}
\ead{E-mail:eric.howell@uwa.edu.au}


\label{firstpage}
\begin{abstract}
{We investigate joint low-latency gravitational wave (GW) detection and prompt electromagnetic (EM) follow-up observations of coalescing binary neutron stars (BNSs). Assuming that BNS mergers are associated with short duration gamma ray bursts (SGRBs), we evaluate if rapid EM follow-ups can capture the prompt emission, early engine activity or reveal any potential by-products such as magnetars or fast radio bursts. To examine the expected performance of extreme low-latency search pipelines, we simulate a population of coalescing BNSs and use these to estimate the detectability and localisation efficiency at different times before merger. Using observational SGRB flux data corrected to the range of the advanced GW interferometric detectors, we determine what EM observations could be achieved from low-frequency radio up to high energy $\gamma$-ray. We show that while challenging, breakthrough multi-messenger science is possible through low latency pipelines.}
\end{abstract}
\graphicspath{{figs/}{./}}

\section{Introduction}
\label{intro}
The first observations by the advanced generation of interferometric gravitational wave (GW) detectors is set to ignite a new era of astronomy in which GW triggered follow-ups by facilities operating outside the GW spectrum will become common place. Complementary observations from electromagnetic (EM), neutrino or high energy particle facilities would not only improve the confidence of a GW detection, but also maximise the science return via a wealth of additional astrophysical information.

The advanced generation of detectors will be led by Advanced LIGO (aLIGO) \cite{aLIGO_2015} during the second half of 2015 and followed by Advanced Virgo (AdV) \cite{AdV_2015} a year later \cite{LSC_Prospects_aLIGO_2013}. Coalescing systems of BNSs are one of the most promising sources of GWs in terms of detection \cite{thorne87};  at design sensitivity (around 2018), the expected detection rates of these sources are of order 0.4 to 400 BNS/year \cite{Abadie2010CQGra,CowardHowellPiran_2012}. One important aspect of these well-modeled inspiralling systems is that a detection can be made tens of seconds before the merger if enough waveform cycles can be detected to boost the signal-to-noise ratio \cite{manzotti12,cannon12}. This could allow an \emph{extreme low-latency} alert to be sent out to EM facilities as near real-time as possible to catch the fading EM signature. Such prompt follow-ups could provide valuable insight into the inner workings of these cataclysmic events.

Here we report on the results of a study conducted by Chu et al. \cite{Chu_inprep} which examined the capability of a range of EM instruments for rapid follow-ups from extreme low-latency GW triggers. We present here only the main results assuming an aLIGO/AdV network at design sensitivity. We refer the interested reader to Chu et al. for a complete analysis including results from larger GW Network configurations.
\begin{figure*}
\includegraphics[scale = 0.1]{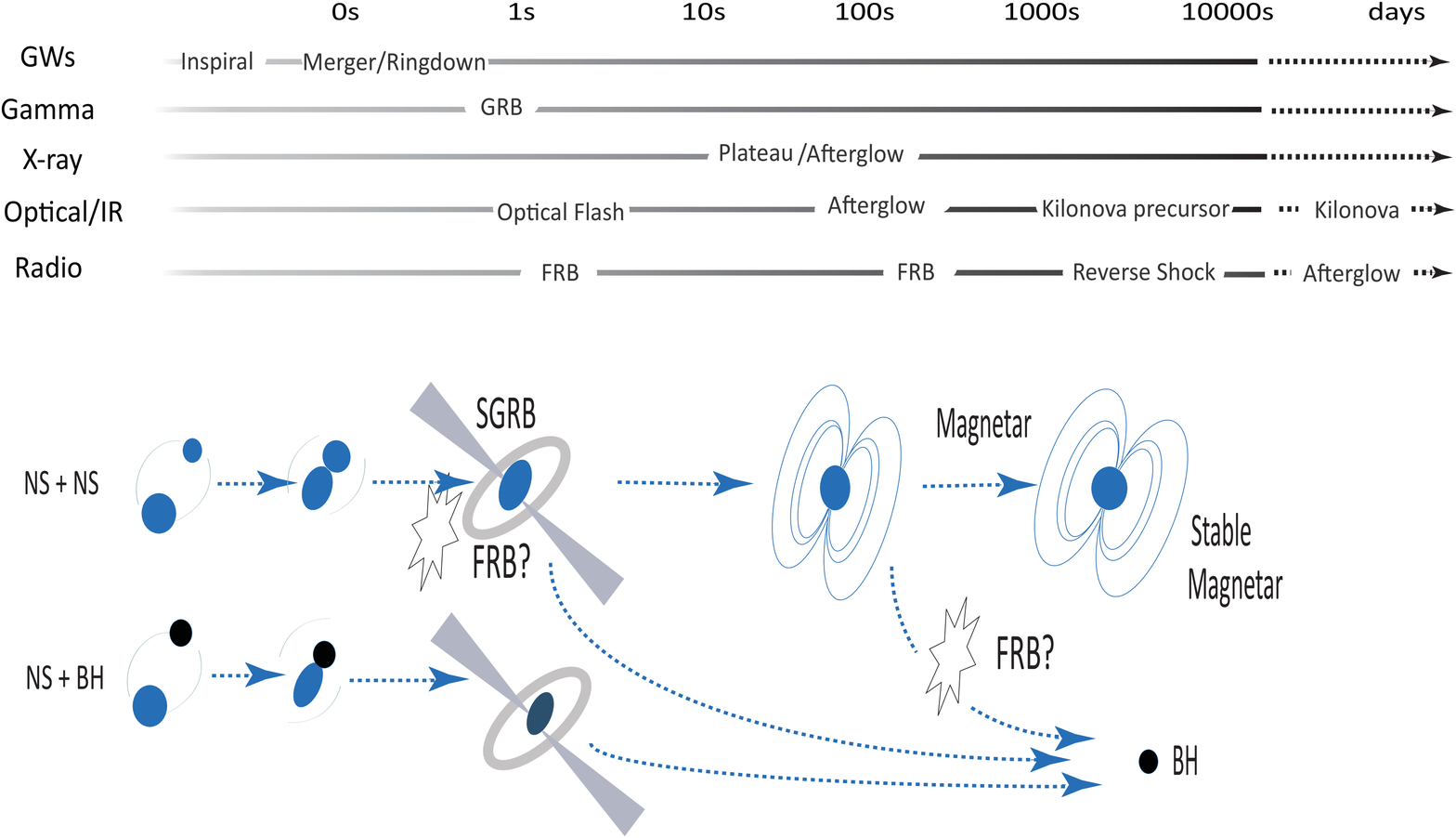}\\
\caption{A cartoon illustrating some of the possible EM pathways, timescales and end products of coalescing systems of NSs and BHs. Adapted from Chu et al. \cite{Chu_inprep}.}
  \label{fig_mm_scenarios}
\end{figure*}
\section{The EM pathways for BNS mergers}

Figure~\ref{fig_mm_scenarios} summarises some of the popular scenarios for the expected EM signatures from BNS mergers. These are based on the proposed link between short duration gamma ray bursts (SGRBs) and the merger of of compact objects \cite{Paczynski1986ApJ,Eichler1989Natur,gehrels05,Berger2005Natur, Bloom2006ApJ}. The predictions are highly uncertain and could lead to various pathways and end products. The anticipated onset time, duration, and observable wavelengths are shown in the figure.

Figure~\ref{fig_mm_scenarios} shows that SGRBs could be accompanied by an optical flash or a fast radio burst (FRB) \cite{Totani2013PASJ}; an FRB could also result from the collapse of a merger product to a black hole \cite{Zhang2014ApJ}. If a stable or supra-massive magnetar is formed, the long lived X-ray plateaus observed in many SGRBs could be the signature of direct dissipation of magnetar wind energy \cite{Rowlinson2010MNRAS,Rowlinson2013MNRAS,Zhang2013ApJ}. At later times, predictions for an optical/IR kilonova \cite{metzger10} are supported though recent observations \cite{Berger2013ApJ,Tanvir2013Natur}. For a relatively small opacity, a kilonova precursor may also appear soon after the merger \cite{Metzger_2014}.

\section{Accessing the low-latency GW detector response and EM follow-up}
Since the initial LIGO era a number of low-latency GW trigger-generation pipelines have been proposed and tested ~\cite{first_low_latency_inspiral, mbta,  cannon12, jing,spiir}. The low latency localization pipeline \verb"BAYESTAR" ~\cite{leo_bayestar,first2years} has shown that localization can be achieved in seconds with potential speed ups proportional to the number of CPU cores. For success, wide-field EM telescopes capable of rapid response within tens of seconds and with fields-of-view (FoV) comparable to the error area of GW source sky directions (of order hundreds of deg$^{2}$) will be critical.
\begin{figure*}
\includegraphics[scale = 0.6]{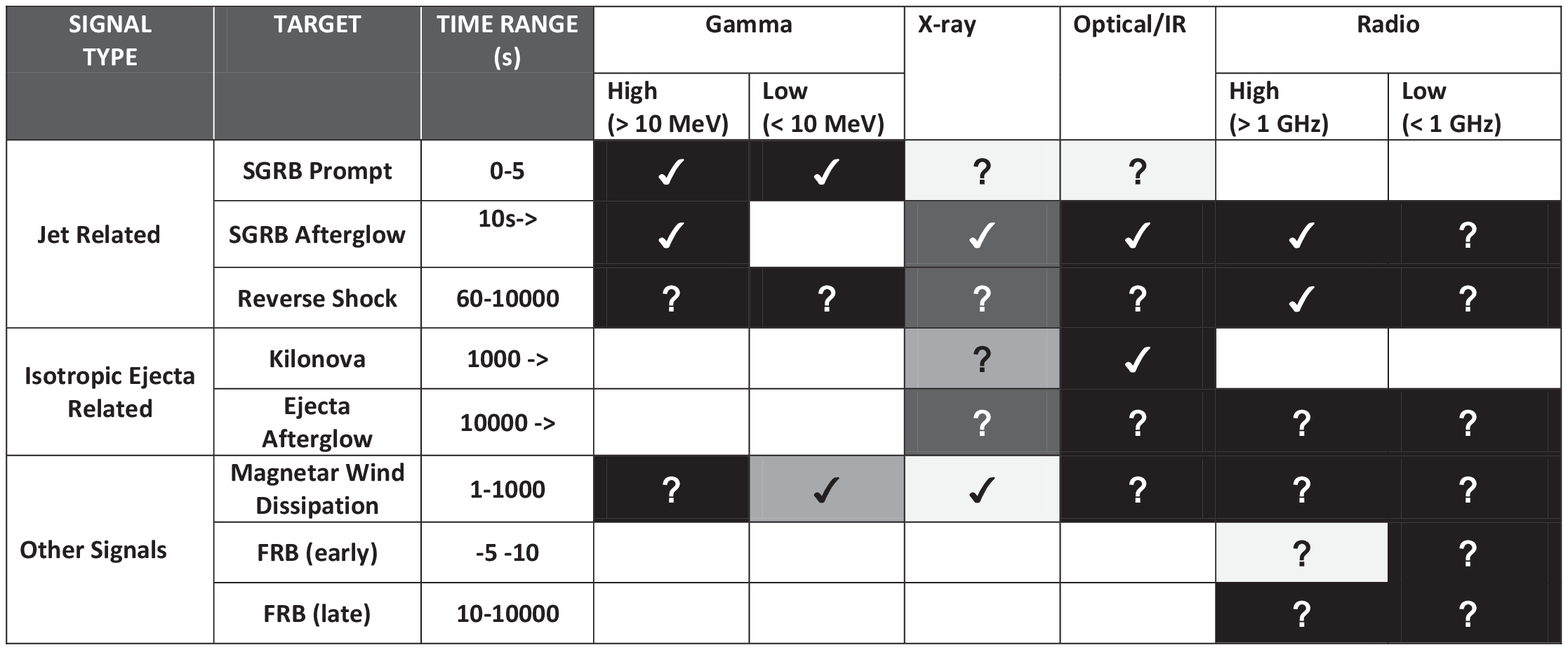}\\
  \caption{Some of the multimessenger observations that could be achieved for our follow-up scenario. The signals are grouped into jetted, isotropic or other (or unknown) signal types; the approximate emission time ranges around merger are also given. As shown by the legend, \textbf{ticks} indicate the EM observations with observational support; \textbf{question marks} indicate those that have been proposed through theoretical models; \textbf{dark cells} indicate there are EM instruments available in the band with sufficiently fast responses, exposure times and FoVs to cover or tile the GW error box within the emission time-range.}
  \label{fig_mm_scenarios1}
\end{figure*}
To test the EM-follow up prospects we simulated a total of 200,000 BNS sources, drawing free parameters of the waveform and the sky location of each event from standard distributions. Around 10$^3$ BNS detections were obtained and the localization accuracy of each was determined using the formalism of \cite{wen10}. Detailed statistics of the detected sample are provided in Chu et al. \cite{Chu_inprep}. For an aLIGO/AdV network the error region required to contain 50\% of the detections, assuming sources are detected 40s before merger, was around 1000 deg$^2$. We assume this estimate, along with a 40s latency to send out the GW trigger.

To explore prompt EM follow-up observations, we estimate the fluxes at different wavelengths by extrapolating data from observed EM emissions associated with SGRBs and convert to 200 Mpc. We then consider the FoVs, sensitivities and response times for a range of EM telescopes to estimate their ability to capture a prompt EM signature.
\section{Results and discussion}
Figure \ref{fig_mm_scenarios1} shows that at low frequency radio (MWA \cite{Tingay2013PASA}; FoV:610\,deg$^{2}$@150\,MHz; response in tens of secs) and both high and low-energy $\gamma$-ray (e.g. Fermi \cite{Meegan_Fermi_2009}; GBM FoV:30000\,deg$^{2}$, LAT FoV:6564\,deg$^{2}$) the large FoVs and the fast response potential of MWA show that post-prompt ($>$ 5s) breakthrough observations could be realised \cite{2015PASA...32...46H}. In the X-ray, small FoV instruments such as \emph{Swift} are hindered by the large GW error region; a requirement for reconstructed sky locations to overlap with nearby galaxies \cite{Evans2012ApJS} could improve this situation.

In the optical, even for a wide field, fast response telecope like ZTF (\cite{Bellm_2014}; FoV:47\,deg$^{2}$; exposure 30s; 20.5-21 mag), to cover the 1000 deg$^{2}$ error region with an exposure times of 30s would require hundreds of seconds. Similar instruments will require sophisticated tiling strategies and would benefit from a co-ordinated observational scheme (see \cite{SingerPTF2015ApJ} for search strategies using the iPTF). A wide FoV X-ray instrument capable of rapid pointing, such as the NASA proposed 900 deg$^{2}$ FoV soft X-ray instrument ISS-Lobster \cite{Camp2013}, would be invaluable for such an approach.

\section*{ACKNOWLEDGMENTS}
 EJH acknowledges support from a UWA Research Fellowship and LW support from the Australian Research Council. We thank C. Berry (LSC review) and the anonymous referee for valued comments that have improved the manuscript.
\vspace{5mm}

\providecommand{\newblock}{}

 \label{lastpage}
\end{document}